\input{aipcheck}

\documentclass[
    ,final            
  ]
  {aipproc}

\layoutstyle{6x9}

\SetInternalRegister\hbadness{8000} 

\begin{document}

\title{Powerful gravitational-wave bursts from supernova neutrino oscillations }


\author{Herman J. Mosquera Cuesta }
{ address={Centro Brasileiro de Pesquisas F\'{\i}sicas \\ 
CEP: 22290-180 Rio de Janeiro, Brazil}, email={hermanjc@cbpf.br}, }

\iftrue
\author{Karen Fiuza}
{address={Instituto de F\'{\i}sica - Universidade Federal do Rio Grande do Sul\\
CEP: 91501-970 Porto Alegre, Brazil}, email={kfiuza@if.ufrgs.br},}
\fi  

\copyrightyear  {2004}

\begin{abstract}
During supernova core collapse and bounce resonant active-to-active, as
well as active-to-sterile, neutrino ($\nu$) oscillations can take
place. Over this phase weak magnetism increases antineutrino mean free
paths, and thus its luminosity.  Because oscillations feed mass-energy
into the target $\nu$ species, the large mass-squared difference
between $\nu$ states implies a huge amount of power to be given off as
gravitational waves (GWs) due to the universal {\it spin-rotation} and
the spin-magnetic coupling driven $\nu$ anisotropic flow, which is coherent
over the oscillation length. The spacetime strain produced is about two
orders of magnitude larger than those from $\nu$ difussion or neutron
star matter anisotropies. GWs observatories as LIGO, VIRGO, GEO-600,
TAMA-300, etc., can search for these bursts far out to the VIRGO
cluster of galaxies.
\end{abstract}

\date{\today}
\maketitle

\section{Generating GWs from neutrino conversions}

$\nu$ oscillations into sterile $\nu$s change dramatically the energy
and momentum (linear and angular) configuration of the proto-neutron
star (PNS). Flavor conversions into sterile $\nu$s drive a large mass
and energy loss from the PNS because once they are produced they freely
escape from the star  \cite{wolfenstein79}. Physically, the potential
$V_s(x)$, for sterile neutrinos in dense matter is zero, and their
probability of reconversion, still inside the star, into active species
is quite small \cite{mosquera00}. The remaining configuration of the
star must also reflect this loss, i.e., the active neutrinos were
coupled to the neutron matter \cite{mosquera02}. Hence its own matter
and energy distribution becomes quadrupolar. Because such a quadrupole
configuration (of matter and energy trapped by the PNS) keeps changing
over the timescale ($\Delta T_{\rm osc} \sim 3-10$~ms after the SN core
bounce) for which most of the oscillations take place, then GWs must
also be emitted from the star over that transient\cite{mosquera-fiuza04}.

In $\nu$ oscillations involving active species the key feature is that
mass and energy is coherently relocated from one region to another
inside the PNS \cite{mosquera-fiuza04}, specially because of the {\it
weak magnetism} of antineutrinos that allows them to have larger mean
free paths \cite{horowitz02}. Besides, the resonance (MSW) condition 
indicates that only ${\nu}_e$s and the antineutrinos $\bar{\nu}_\mu$ 
and $\bar{\nu}_\tau$ can undergo resonant conversions \cite{mosquera00}.  
Thus neutral-current interacting $\nu$ species must be the very first constituents of the $\nu$ burst from any supernova since most $\nu_e$s are essentially trapped.

\section{Anisotropic neutrino outflow}

\begin{figure}
\label{4-dupole}
 \includegraphics[height=.25\textheight]{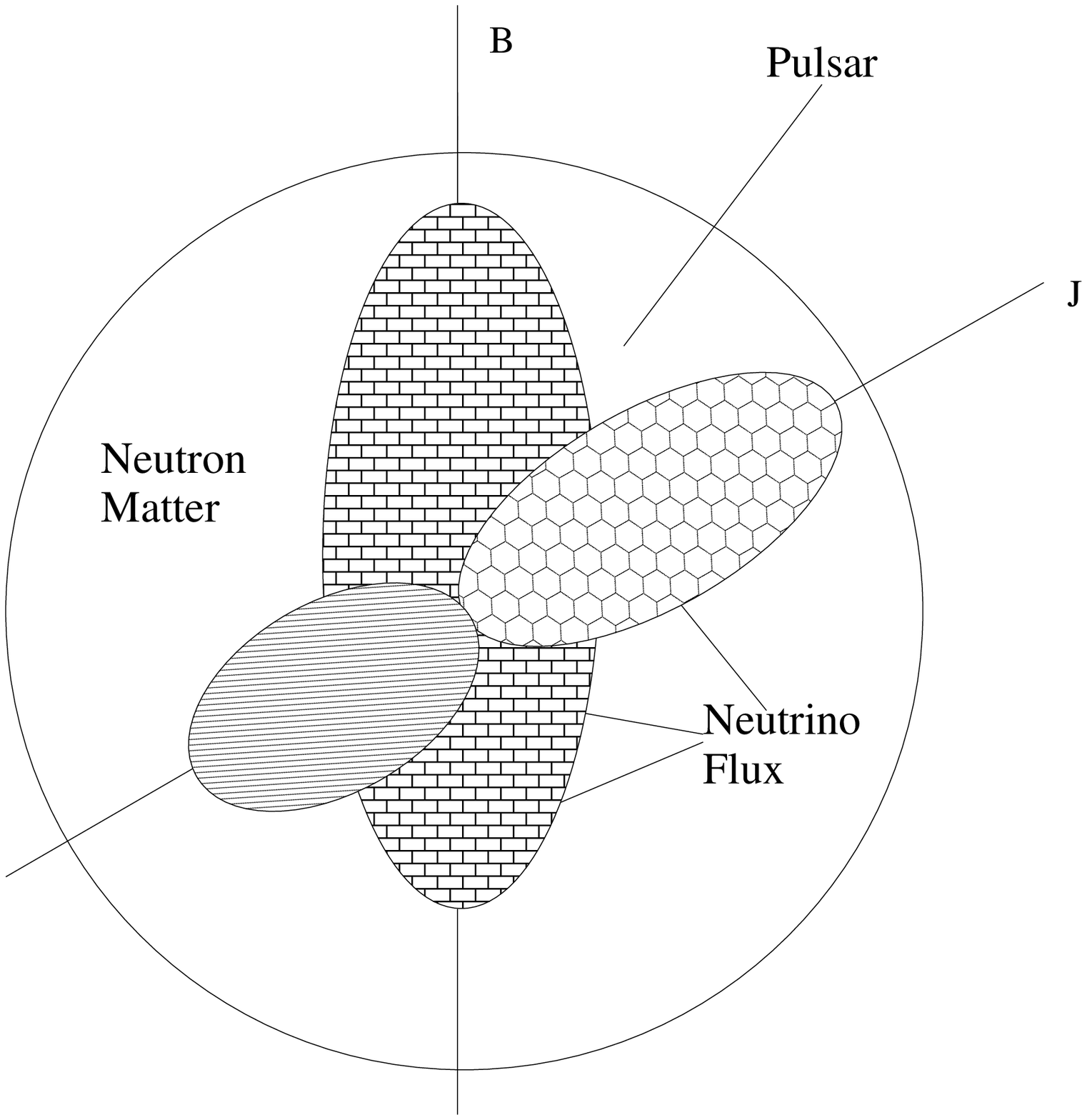}
\caption{Schematic representation of the $\nu$-flux distribution (hatched regions) in a nascent pulsar. The at least quadrupolar distribution is evident.}
\end{figure}

The $\nu$ {\it asymmetry parameter} $\alpha$ \cite{mosquera-fiuza04} 
measures how large is the deviation of the $\nu$-flux from a spherical 
one. As previously discussed \cite{mosquera-fiuza04,loveridge}, once a 
PNS is forming following the SN collapse both the $\nu$-spin coupling to 
rotation and the $\nu$-spin coupling to the magnetic ($\vec{B}$) field 
drives an effective momentum, and thus energy flux, asymmetry along both 
the axes of angular momentum ($\vec{J}$) and $\vec{B}$-field (see Fig \ref{4-dupole}) (notice in passing that the anisotropy origin in a our 
approach is alike from that in Ref. \cite{loveridge}). The combined 
action on the escaping neutrinos of a magnetized and rotating background 
spacetime makes their $\nu$-sphere a decidedly distorted surface, with 
at least a quadrupolar distribution. The $\nu$ flow anisotropy can be 
estimated in a novel and self-consistent fashion by defining $\alpha$ as 
the ratio of the total volume filled by the distorted $\nu$-spheres to 
that of the "spherical" PNS \cite{mosquera-fiuza04}. For a PNS 
\cite{burrows95} with parameters given in our original paper \cite{mosquera-fiuza04} one obtains \cite{burrows96}
\begin{equation}
 \alpha_{\rm min} = \frac{ V_{ellips.} + V_{lemnisc.} }{ V_{\rm PNS} } \sim 0.11-0.01\; . \label{param-alpha}
\end{equation}

This is a very new result. The definition in Eq.(\ref{param-alpha}) does take into account 
all of the physics of the $\nu$ oscillations: luminosity, density 
gradients and angular propagation \cite{burrows95,muller97}.

\section{GWs energetics from $\nu$ luminosity}

If $\nu$ oscillations do take place in the SN core, then the most
likely detectable GWs signal should be produced over the time interval
for which the conditions for flavor conversions to occur are kept,
i.e., $\sim (10^{-1}-10)$~ms. This timescale implies GWs frequencies
in the interval: $[10 \leq f_{\rm{GWs}} \leq 0.1]$~kHz, centered at 1~kHz,
because of the maximum $\nu$ production around $1-3$~ms after core
bounce \cite{mosquera00}. This frequency range includes the optimal
bandwidth for detection by ground-based observatories. For a 1~ms
conversion time span the $\nu$ luminosity reads
\begin{equation}  
L_{\nu}  \equiv   \frac{\Delta  E_{\nu_a  \longrightarrow  \nu_s}
 }{\Delta  T_{\textrm{osc}}  }  \sim  \frac{3\times 10^{51}  \rm  erg}
 {1\times  10^{-3} \rm  s}  = 3  \times  10^{54} \frac{\rm  {erg}}{\rm
 s}\; . 
\label{luminosity}
\end{equation}

The GWs burst amplitude of the signal generated by the non-spherical outgoing front of oscillation-created $\nu_s$ (see details in \cite{mosquera00,mosquera02,mosquera-fiuza04}) can be written as
\begin{equation}
 h_\nu  =  \frac{2G}{c^4  R}  \left(\Delta t  \times L_\nu  \times
\alpha\right) \simeq  |A| \left[\frac{55\rm kpc}{R}\right]
\left(\frac{\Delta  T   }{10^{-3}  \rm  s}\right)  \left[\frac{L_\nu}{
3\times     10^{54}\frac{\rm     erg}     {\rm    s}     }     \right]
\left(\frac{\alpha}{0.1}\right) \; , \nonumber \label{burrows96a}
\end{equation}
where $|A| = 4 \times 10^{-23}~ {\rm Hz}^{-1/2}$ is the amplitude. From 
this equation, and the definition of GWs flux, one can derive the GWs 
luminosity as
\begin{equation} 
L_{\textrm{GWs}} \sim 10^{48-50} ~~ \frac{\rm {erg}}{\rm s}
~~  \left[\frac{L_\nu}{3\times  10^{54} \frac{\rm {erg}}{\rm  s}}
\right]^2 \left(\frac{\alpha}{0.1-0.01}\right)^2\;.
\end{equation}

This power is about $10^{5-3} \times L^{\rm Diff.}_\nu$ the luminosity 
from $\nu$ diffusion inside the PNS \cite{mosquera02}. 

\section{Conclusion} 

The GW signal from the process here outlined is expected to irradiate much more energy than current mechanisms figured out to drive the NS dynamics at birth do. A luminosity this large (Eq.(\ref{luminosity})) would turn these bursts the detectable strongest GW signal from any
SN. In proviso, we argue that a GW pulse that strong could  have been detected during SN1987a from the Tarantula Nebula in the Milky Way's satellite galaxy Large Magellanic Cloud, despite the low sensitivity of detectors at the epoch. In such a case, the GW burst must have been
correlated in time with the earliest arriving $\nu$ burst constituted of some active species given off during the very early oscillation transient where some $\nu_e$s went into $\nu_{\mu}$s, $\nu_{\tau}$s or $\nu_s$s, this last in principle undetectable. Thenceforth, it could be of worth to reanalyze the data collected from that event by taking careful follow-up of their arrival times, if appropriate timing was available at that moment.

\begin{theacknowledgments}
KF thanks Sandra D. Prado. HJMC acknowledges support from FAPERJ (Brazil).
\end{theacknowledgments}


\bibliographystyle{aipproc}   


\bibliography{hadron}

\IfFileExists{\jobname.bbl}{}
 {\typeout{}
  \typeout{******************************************}
  \typeout{** Please run "bibtex \jobname" to optain}
  \typeout{** the bibliography and then re-run LaTeX}
  \typeout{** twice to fix the references!}
  \typeout{******************************************}
  \typeout{}
 }

\end{document}